# Deterministic Black-Box Identity Testing $\pi$-Ordered Algebraic Branching Programs

Maurice Jansen[*]    Youming Qiao[*]    Jayalal Sarma M.N.[*]

November 15, 2018

**Abstract**

In this paper we study algebraic branching programs (ABPs) with restrictions on the order and the number of reads of variables in the program. An ABP is given by a layered directed acyclic graph with source $s$ and sink $t$, whose edges are labeled by variables taken from the set $\{x_1, x_2, \ldots, x_n\}$ or field constants. It computes the sum of weights of all paths from $s$ to $t$, where the weight of a path is defined as the product of edge-labels on the path. Given a permutation $\pi$ of the $n$ variables, for a $\pi$-*ordered* ABP ($\pi$-OABP), for any directed path $p$ from $s$ to $t$, a variable can appear at most once on $p$, and the order in which variables appear on $p$ must respect $\pi$. An ABP $A$ is said to be of read $r$, if any variable appears at most $r$ times in $A$.

Our main result pertains to the identity testing problem, i.e. the problem of deciding whether a given $n$-variate polynomial is identical to the zero polynomial or not. Over any field $\mathbb{F}$ and in the black-box model, i.e. given only query access to the polynomial, we have the following result: read $r$ $\pi$-OABP computable polynomials can be tested in DTIME$[2^{O(r \log r \cdot \log^2 n \log \log n)}]$. In case $\mathbb{F}$ is a finite field, the above time bound holds provided the identity testing algorithm is allowed to make queries to extension fields of $\mathbb{F}$.

Our next set of results investigates the computational limitations of OABPs. It is shown that any OABP computing the determinant or permanent requires size $\Omega(2^n/n)$ and read $\Omega(2^n/n^2)$. We give a multilinear polynomial $p$ in $2n+1$ variables over some specifically selected field $\mathbb{G}$, such that any OABP computing $p$ must read some variable at least $2^n$ times. We prove a strict separation for the computational power of read $(r-1)$ and read $r$ OABPs. Namely, we show that the elementary symmetric polynomial of degree $r$ in $n$ variables can be computed by a size $O(rn)$ read $r$ OABP, but not by a read $(r-1)$ OABP, for any $0 < 2r-1 \le n$. Finally, we give an example of a polynomial $p$ and two variables orders $\pi \ne \pi'$, such that $p$ can be computed by a read-once $\pi$-OABP, but where any $\pi'$-OABP computing $p$ must read some variable at least $2^n$ times.

## 1 Introduction

The polynomial identity testing problem (PIT) is the question of deciding, given an arithmetic circuit $C$ with input variables $x_1, x_2 \ldots x_n$ over some field $\mathbb{F}$, whether the polynomial computed by $C$ is identical to the zero polynomial in the ring $\mathbb{F}[x_1, x_2, \ldots x_n]$. Efficient algorithms for PIT are important both in theory and in practice. Randomized algorithms were given independently by Schwartz [1] and Zippel [2].

Finding *deterministic* algorithms for PIT plays a crucial role in computational complexity theory. Kabanets and Implagliazzo [3] showed that giving a deterministic subexponential time algorithm for PIT implies that either NEXP $\not\subseteq$ P/*poly*, or that the permanent has no *poly*-size arithmetic circuits. Agrawal [4] showed that giving a deterministic *black-box* algorithm for PIT yields an explicit multilinear polynomial that has no subexponential size arithmetic circuits. In [4] a program was outlined explaining how making progress towards the latter kind of algorithm for PIT has the potential of resolving Valiant's Hypothesis, which states that the algebraic complexity classes VP and VNP are distinct.

[*]Institute for Theoretical Computer Science, Tsinghua University, Beijing, China. maurice.julien.jansen@gmail.com, jimmyqiao86@gmail.com, jayalal@tsinghua.edu.cn. This work was supported in part by the National Natural Science Foundation of China Grant 60553001, and the National Basic Research Program of China Grant 2007CB807900,2007CB807901.

For optimists certainly, the situation is tantalizing, as Agrawal and Vinay [5] showed that the black-box derandomization of PIT for only depth-4 circuits would yield a nearly complete derandomization for general arithmetic circuits. Recent progress on the PIT problem has been impressive. See [6] for a recent survey.

In this paper, we contribute to the above mentioned lower bounds program by considering black-box identity testing *ordered algebraic branching programs* (OABPs), which where introduced in [7]. Algebraic branching programs have computational power somewhere in between arithmetic formulas and circuits. The OABP is the arithmetic analogue of the *ordered binary decision diagram* (OBDD), which was introduced by Bryant [8]. Some polynomials can be succinctly represented in the OABP model. For example, we show that the elementary symmetric polynomial of degree $k$ in $n$ variables can be elegantly described by a grid shaped OABP of size $O(kn)$. As our lower bounds show, a succinct OABP-representation is not available for every polynomial. The situation is similar to what is well-known for OBDDs. In practice this may be outweighed by the fact that PIT can be solved efficiently for OABPs. As argued by Raz and Shpilka [9], part of the popularity of OBDDs can be explained by the fact that identity testing (and hence equivalence testing) can be done efficiently for the model.

In [9] a polynomial-time non-black-box algorithm was given for identity testing non-commutative formulas, and more generally non-commutative ABPs. Identity testing OABPs reduces to PIT for non-commutative ABPs, and hence can be done *non-black-box* in polynomial time. Namely, if we take an OABP $A$ computing some polynomial $f$ over commuting variables, and if we let $f'$ be the evaluation of $A$, where we restrict the variables to be non-commuting, then it can be observed that $f \equiv 0 \Leftrightarrow f' \equiv 0$. Giving a *black-box* algorithm for testing non-commutative formulas and ABPs is currently a major open problem. Our main result implies we have a DTIME$[2^{O(polylog(n))}]$ black-box algorithm for testing OABPs with $polylog(n)$ many reads.

We remark that any *read-once formula* (ROF) can be simulated by an OABP. Black-box Identity testing sum-of-$k$ ROFs was studied in [10], and this was subsequently generalized to the sum-of-$k$ read-once ABPs in [11]. These results show the difficulty of making generalizations in this area to models beyond read-once. For example, by [11] we have an $n^{O(\log n)}$ black-box test for sum-of-two read-once ABPs, but for testing a single read-twice ABP, currently nothing is known beyond brute-force methods. Our result is significant, in that the techniques apply to a model where the multiple reads take place within one *monolithic* ABP. This opens up a new thread of progress in the direction of identity testing unrestricted ABPs. We refer to [12] for a direct connection between this, and proving lower bounds for the *determinantal complexity* of explicit polynomials. The latter is what the separation of VP and VNP requires.

Finally, we mention the connection to PIT for multilinear formulas raised in [13]. Our results can be applied to black-box identity testing "ordered multilinear formulas" with few reads (say $polylog(n)$). The latter can be defined for any given variable order $\pi$, by requiring that for each multiplication gate $g = g_1 \times g_2$ in the formula, variables in the subformula rooted at $g_1$ should either all be smaller or all be larger w.r.t. $\pi$ than variables in the subformula rooted at $g_2$. By applying the construction of [14], judiciously to keep the order, a formula of this kind can be simulated by a $\pi$-OABP. This then gives another important special case of PIT for multilinear formulas for which a black-box algorithm is known (the other case being sum-of-$k$ ROFs).

## 1.1 Techniques

Towards the identity testing algorithm, the first step is to show that, without increasing the number of reads, any $\pi$-OABP can be made *oblivious*, i.e. all variables in a layer must be identical. Then we construct a generator $\mathcal{G}(z)$ for $\pi$-oblivious ABPs. This is a polynomial mapping $\mathbb{F}^\ell \to \mathbb{F}^n$ such that for any $f \in \mathbb{F}[x_1, \ldots, x_n]$ computed by a $\pi$-oblivious ABP, $f \equiv 0 \Leftrightarrow f(\mathcal{G}) \equiv 0$. From this, one obtains an efficient black-box test, provided the number of $z$-variables $\ell$ and the degree of $\mathcal{G}$ is "small".

For illustrative purposes, let us consider an oblivious ABP $A$ with variable order $x_1, x_2, \ldots, x_{2n}$ of small width $w$, rather than small number of reads, and suppose it computes $f \not\equiv 0$. In order to achieve $\ell = O(w \log n)$, we cut $A$ in the middle layer. This gives a decomposition (say) $f =$ $\sum_{i \in [w]} g_i(x_1, \ldots, x_n) h_i(x_{n+1}, \ldots, x_{2n})$. Then we want $f(\mathcal{G}) = \sum_{i \in [w]} g_i(\mathcal{G}_1, \ldots, \mathcal{G}_n) h_i(\mathcal{G}_{n+1}, \ldots, \mathcal{G}_{2n}) \not\equiv 0$. We would like to use recursion on the $g_i$s and $h_i$s, but in order to get $\ell$ small, this means $\mathcal{G}^u := (\mathcal{G}_1, \ldots, \mathcal{G}_n)$ and $\mathcal{G}^d := (\mathcal{G}_{n+1}, \ldots, \mathcal{G}_{2n})$ will share most of the variables. This open up the possibility of all kinds of unwanted cancellations to start occurring, resulting in $f(\mathcal{G}) \equiv 0$. However, we do have something going for ourselves, which is that $\{g_i(\mathcal{G}^u)\}_{i \in [w]}$ must "communicate" through a small dimensional space $\mathbb{F}^w$. This allows one to take $\mathcal{G}^d$ identical to $\mathcal{G}^u$, except for an additional component to the input that inflates the dimension of any non-empty finite union of *affine varieties*[1], given by the preimage of a single point in $\mathbb{F}^w$. More or less, $\mathcal{G}(z, z')$ will look like $\mathcal{G}^u(z); \mathcal{G}^d(z, z')$, with $\mathcal{G}^d(z, z') = \mathcal{G}^u(z + T(z'))$, where $T$ is a mapping of $O(w)$ many variables that contains any $w$-dimensional coordinate subspace. Doing so, we only add $O(w)$ many variables per inductive step. This mirrors the pseudorandom generator construction of Impagliazzo, Nisan and Wigderson [16] in the Boolean world. To make an analogy, $z + T(z')$ can be thought of as similar to taking a vertex (we pick $z$) and adjacent edge (we move by $T(z)$) on an expander graph.

Necessarily, our final construction will be more complicated than the above sketch, since we assume a bound on the number of reads instead of the width. This will be dealt with by taking a partial derivatives w.r.t. a centrally local variable $x_k$ in the ABP. Taking the derivative w.r.t. $x_k$ has the net effect of cutting down the width of the $x_k$-layer of $A$.

## 1.2 Organization

The rest of this paper is structured as follows. Section 2 contains preliminaries. Section 3 contains a proof of a "decomposition lemma" we require for Section 4. There we do the core of the technical work by giving the generator for $\pi$-oblivious ABPs. Then we apply this generator in Section 5 to give the black-box PIT algorithm for $\pi$-OABPs. Finally, we prove the results regarding computational limitations of OABPs mentioned in the abstract in Section 6.

## 2 Preliminaries

For a natural number $n$, we denote the set $\{1, 2, \ldots, n\}$ by $[n]$. For an $n$-tuple $a = (a_1, a_2, \ldots, a_n)$ and $m$-tuple $b = (b_1, b_2, \ldots, b_m)$, we denote its concatenation $(a_1, a_2, \ldots, a_n, b_1, b_2, \ldots, b_m)$ by $a \# b$. Let $X = \{x_1, x_2, \ldots, x_n\}$ be a set of variables and let $\mathbb{F}$ be a field. For a polynomial $f \in R := \mathbb{F}[X]$, if it is identical to the zero polynomial of the ring $R$, we write $f \equiv 0$. If the degree of any variable of $f$ is bounded by one, $f$ is said to be *multilinear* (even if $f$ has a constant term). We say $f$ *depends on* $x_i$, if the formal partial derivative $\partial f / \partial x_i \not\equiv 0$. $Var(f)$ denotes the sets of variables $f$ depends on. For a set of polynomial $f_1, \ldots, f_m \in \mathbb{F}[X]$, we say that they are *independent* if for all $a \in \mathbb{F}^m$ with $a \neq \overline{0}$, $\sum_{i \in [m]} a_i f_i \not\equiv 0$.

### 2.1 Computational Models

We import the following definition and subsequent notations from [11]:

**Definition 1.** *An algebraic branching program is a 4-tuple $A = (G, w, s, t)$, where $G = (V, E)$ is an edge-labeled directed acyclic graph for which the vertex set $V$ can be partitioned into levels $L_0, L_1, \ldots, L_d$, where $L_0 = s$ and $L_d = t$. Vertices $s$ and $t$ are called the source and sink of $A$, respectively. Edges may only go between consecutive levels $L_i$ and $L_{i+1}$.*

*The label function $w : E \to X \cup \mathbb{F}$ assigns variables or field constants to the edges of $G$. For a path $p$ in $G$, we extend the weight function by $w(p) = \prod_{e \in p} w(e)$. Let $P_{i,j}$ denote the collection of all paths $p$ from $i$ to $j$ in $G$. The program $A$ computes the polynomial $\sum_{p \in P_{s,t}} w(p)$. The size of $A$ is taken to be $|V|$, and the read of $A$ is the maximum of $|w^{-1}(x_i)|$, over all $x_i$'s. The depth of $A$ equals $d$, and the width of $A$ is the maximum of $|L_i|$, over all $L_i$'s.*

---
[1] Keeping with the terminology in [15], an *algebraic set* is the set of common zeroes of a list of polynomials. Affine varieties are algebraic sets, which are *irreducible* in the *Zariski-topology* (See Appendix B).

Algebraic branching programs where first introduced by Nisan [17]. Our definition differs in the respect that [17] requires edge labels to be linear forms. We use the following notation: for an arc $e = (v, w)$ in ABP $A$, $begin(e) = v$ and $end(e) = w$. We let $source(A)$ and $sink(A)$ stand for the source and sink of $A$. For any nodes $v, w$ in $A$, we denote the subprogram with source $v$ and sink $w$ by $A_{v,w}$. We use $\widehat{A}$ to denote the polynomial computed by $A$, and in particular, $\widehat{A_{v,w}}$ is the polynomial computed by the subprogram $A_{v,w}$. A *layer* of an ABP $A$ is the subgraph induced by two consecutive levels $L_i$ and $L_{i+1}$ in $A$.

**Definition 2.** *An ABP $A$ is called a read $r$ ABP, if its read is bounded by $r$. We also denote this by saying that $A$ is an $R_r$-ABP. A polynomial $f \in \mathbb{F}[X]$ is called an $R_r$-ABP-polynomial if there exists a $R_r$-ABP computing $f$.*

**Definition 3.** *Let $\pi$ be a permutation of $[n]$. An ABP $A$ is $\pi$-ordered, if on every directed path $p$ in $A$, if a variable $x_i$ appears before $x_j$ on $p$, then $\pi(i) < \pi(j)$. For an ABP $A$ we say it is ordered if it is $\pi$-ordered w.r.t. some permutation $\pi$.*

For a $\pi$-ordered ABP ($\pi$-OABP) variables appear (with possibly omissions) on any path from source to sink in the order $x_{\pi^{-1}(1)}, x_{\pi^{-1}(2)}, \ldots, x_{\pi^{-1}(n)}$. We will speak of the latter sequence as the *variable order* of $A$. Ordered algebraic branching programs where first studied in [7], but with respect to the homogeneous ABP definition of [17]. There the ordering condition states that on any path $p$, for any edge $e_1$ appearing before $e_2$ on $p$, if $e_1$ is labeled by $\sum_{i \in [n]} a_i x_i$, and $e_2$ is labeled by $\sum_{i \in [n]} b_i x_i$, then all variables in $\{x_i : a_i \neq 0\}$ appear before all variables in $\{x_i : b_i \neq 0\}$ in the variable order. The usual "homogenization trick" of splitting nodes into parts computing homogeneous components can be used to convert any OABP to the model of [7] (one also needs to collapse circuitry going over constant wires). This outlines a proof of the second part of the following lemma (the first part being obvious):

**Lemma 1.** *For any permutation $\pi$ of $[n]$ we have the following:*

1. *A homogeneous $\pi$-ordered ABP of size $s$ with linear forms as edge labels can be converted into an equivalent $\pi$-OABP with weight function $w : E \to X \cup \mathbb{F}$ of size $O(ns)$.*

2. *For any $\pi$-OABP of size $s$ computing a homogeneous polynomial of degree $d$, there exists an equivalent homogeneous $\pi$-ordered ABP of size $O(sd)$ with linear forms as edge labels.*

An ABP is called *oblivious*, if for any layer all variables are the same. We call a layer an *$x$-layer*, if $x$ labels some of the edges in that layer, for $x \in X$. Layers with variables are called *variable layers*. Layers without variables are called *constant layers*. We say an ABP is $\pi$-*oblivous*, if it is $\pi$-ordered and oblivious. We have the following lemma for converting any $\pi$-OABP into a $\pi$-oblivious ABP. Note the lemma preserves read.

**Lemma 2.** *For any permutation $\pi$ of $[n]$, given a $\pi$-OABP $A$ over $n$ variables of size $s$ and read $r$, there is an equivalent $\pi$-oblivious ABP $B$ of size $O(sn)$, width $\leq 2s$ and read $r$.*

*Proof.* Wlog. we assume that $\pi$ is the identity permutation. In $B$, the source will be followed by levels with node sets $V_1, W_1, V_2, W_2, \ldots, V_n, W_n, V_{n+1}$ (in this very order). We will grant ourselves the convenience of also having constant labeled wires that go from $V_i$ to $V_{i+1}$ directly, thus skipping $W_i$. Obviously, this can be fixed by at most doubling the width of the ABP and without any increase in the number of reads. For any $i$, nodes in $V_i$ will be in one-to-one correspondence with nodes in $A$. Namely, for each node $v$ in $A$, we allocate a node $v_i$ in $V_i$. Similarly for $W_i$.

Let $s = source(A), t = sink(A)$ and $\overline{s} = source(B)$. We will arrange to following two properties to hold. Firstly, for any $i \in [n+1]$, for any $v_i \in V_i$,

$$\widehat{B_{\overline{s},v_i}} = (A[x_i = \widehat{x_{i+1} = \ldots} x_n = 0])_{s,v}. \tag{1}$$

In other words, the polynomial computed by the subprogram of $B$ with source $\overline{s}$ and sink $v_i$ equals the sum of weights of all paths from $s$ to $v$ in $A$, going over edges that are not labeled with variables from $\{x_i, x_{i+1}, \ldots, x_n\}$.

Secondly, for any $i \in [n]$, for any $w_i \in W_i$, we will arrange that $\widehat{B}_{\overline{s},w_i}$ will equal the sum of weights of all paths $p$ from $s$ to $w$ in $A$ going over edges that are not labeled with variables from $\{x_{i+1}, \ldots, x_n\}$, and with $x_i$ appearing on the last edge of $p$.

We now explain how to connect the wires in $B$. First of all, for any $v_1 \in V_1$, $(A[x_1 = \widehat{x_2 = \ldots x_n} = 0])_{s,v}$ is a constant. So we can establish (1) for $V_1$ by drawing edges to nodes in $V_1$ with appropriately labeled constants.

For any $i \geq 1$, given that wires have been drawn to $V_i$, we draw wires to $W_i$ as follows. For $v_i \in V_i$ and $w_i \in W_i$, we draw a wire $(v_i, w_i)$ with label $x_i$ if and only if $(v, w)$ is an edge in $A$ with label $x_i$. This clearly establishes the desired property for the set $W_i$, provided (1) holds for $V_i$. It is also immediately clear with this definition that the number of reads of $B$ equals the number of reads of $A$.

For $i > 1$, given that wires have been drawn to $W_{i-1}$, we draw wires to $V_i$ as follows. For any node $v$ in $A$, the set $S$ of paths from $s$ to $v$ in $A$ not using edges with labels $\{x_i, x_{i+1}, \ldots, x_n\}$ can be partitioned into $S_1 \cup S_2$, where $S_1$ is the set of paths from $s$ to $v$ that do not use $x_{i-1}$-labeled edges, and $S_2 = S \setminus S_1$. For any path $p \in S_2$, we can write it as $p = p_1 \# p2$, where $p_1$ ends with an $x_{i-1}$-labeled edge, and $p_2$ consists of constant labeled edges only. Hence, for a node $v_i \in V_i$, $(A[x_i = \widehat{x_{i+1} = \ldots x_n} = 0])_{s,v}$ can be computed as the sum of $(A[\widehat{x_{i-1} = x_i = x_{i+1}} = \ldots x_n = 0])_{s,v}$ plus a linear combinations of the weight of all paths $p$ from $s$ to some node $w$ in $A$ going over edges that are not labeled with variables from $\{x_i, \ldots, x_n\}$, and with $x_{i-1}$ appearing on the last edge of $p$. This can be done by simply drawing a wire from node $v_{i-1} \in V_{i-1}$ with label one, and appropriate constant labeled wires from nodes in $W_{i-1}$.

By the above construction, we get $\widehat{B}_{\overline{s}, t_{n+1}} = \widehat{A}_{s,t}$. After removing redundant nodes, splitting wires that skip $W_i$'s, it is clear we have obtained a leveled oblivious ABP with read $r$, size $O(sn)$ and width $\leq 2s$. □

## 2.2 Partial Derivatives Matrix

The following notions are taken from [13]. Let $Y = \{y_1, \ldots, y_n\}$ and $Z = \{z_1, \ldots, z_n\}$. Then for a multilinear polynomial $p(Y, Z) \in \mathbb{F}[Y, Z]$, we can construct the *partial derivatives matrix* $M_p$ w.r.t. $Y$ and $Z$ as follows: rows and columns of the matrix are indexed by $\{0,1\}^n$, and $M_p(e, f)$ is the coefficient of $\prod_{i \in [n]} y_i^{e_i} \prod_{j \in [n]} z_j^{f_j}$, for $e, j \in \{0,1\}^n$. Thus $M_p$ is a $2^n \times 2^n$ matrix. We are mostly interested in the rank of partial derivative matrices, and some of the useful properties are summarized in the following lemma:

**Lemma 3** ([18]). *Given three multilinear polynomials $p, q, r \in \mathbb{F}[Y, Z]$, we have that 1) If $r = p + q$, then $M_r = M_p + M_q$, 2) If $r = p \cdot q$, and $Var(p)$ and $Var(q)$ are disjoint, then $\mathrm{rank}(M_r) = \mathrm{rank}(M_p) \cdot \mathrm{rank}(M_q)$, 3) If $r = p \cdot q$, and $p \in \mathbb{F}[Y]$ and $q \in \mathbb{F}[Z]$, then $\mathrm{rank}(M_r) = 1$.*

In the above, for the first two properties, see Proposition 3.1 and 3.2 in [18]. The third property follows from the second property, by observing that $M_p$ (resp. $M_q$) has only one column (resp. row) that contains non-zero entries.

Given $X = \{x_1, \ldots, x_{2n}\}$, a *partition* $A$ of $X$ is a one-to-one mapping from $X$ to $Y \cup Z$. For a polynomial $p \in \mathbb{F}[X]$, denote by $p^A$ the polynomial obtained by substituting $x \in X$ with $A(x) \in Y \cup Z$. So $p^A \in \mathbb{F}[Y, Z]$. A multilinear polynomial $p \in \mathbb{F}[X]$ is said to be of *full rank* if for any partition $A : X \to Y \cup Z$, $M_{p^A}$ is of full rank. By extending the construction in [19], we show the existence of a polynomial $p$, such that $\partial p / \partial x$ is a full-rank polynomial for every variable $x$ in $p$. A proof of the following theorem has been put in Appendix A.

**Theorem 1.** *Let $n \in \mathbb{N}$ be an integer, and $X = \{x_1, \ldots, x_{2n+1}\}$, $\mathcal{W} = \{w_{i,j,k}\}_{i,j,k \in [2n+1]}$ be two sets of variables. Let $\mathbb{G} = \mathbb{F}(\mathcal{W})$ be the field of rational functions over $\mathbb{F}$ with variables in $\mathcal{W}$. Then there exists an explicit multilinear polynomial $p \in \mathbb{G}[X]$, such that for any $k \in [2n+1]$, $\partial p / \partial x_k$ is of full rank, when working over $X \setminus \{x_k\}$.*

## 2.3 Algebraic Geometry

Any subset $X \subseteq \mathbb{F}^n$ which is the set of simultaneous zeroes of a set of polynomials $f_1, \ldots, f_t \in \mathbb{F}[x_1, \ldots, x_n]$ is called an *algebraic set*. For basic definitions we refer to [20, 15]. For convenience, all theorems we use have been listed in Appendix B. If $X$ and $Y$ are algebraic sets in $\mathbb{F}^n$, we denote by $X + Y$ the subset $\{x + y \in F^n : x \in X, y \in Y\}$. Note that $X + Y$ may not be an algebraic set. We denote by $\overline{X + Y}$ the closure of $X + Y$ is the Zariski-topology (See Appendix B). We need the following two lemmas:

**Lemma 4.** *Let $X \subset \mathbb{F}^n$ be an algebraic set of dimension $0 \leq r < n$. Then for some $(n-r)$-dimensional coordinate subspace $C \subset \mathbb{F}^n$, $\overline{X + C} = \mathbb{F}^n$.*

*Proof.* For a coordinate subspace $C$ denote the canonical projection to $C$ by $\pi_C$. Consider $K = \{0\}^r \times \mathbb{F}^{n-r}$ and $L = \mathbb{F}^r$, which we think of as the complement of $K$ corresponding to the first $r$ coordinates. We have the following two properties: 1) The set $X + K$ equals $\pi_L(X) \times \mathbb{F}^{n-r}$, and 2) $\overline{\pi_L(X) \times \mathbb{F}^{n-r}} = \overline{\pi_L(X)} \times \mathbb{F}^{n-r}$.

By this, $\dim \overline{X + K} = n - r + \dim \overline{\pi_L(X)}$. More generally, it can be seen (by applying isomorphisms to $\mathbb{F}^n$, where we permute the indices), that for any $(n - r)$-dimensional coordinate subspace $C$ with $r$-dimensional complement $D$, $\dim \overline{X + C} = n - r + \dim \overline{\pi_D(X)}$. Hence the lemma follows from the fact that for any $r$-dimensional affine variety there exists a projection $\tau$ to some $r$-dimensional coordinate subspace $E$ such that $\tau(X)$ is dense in $E$, i.e. $\dim \overline{\pi_D(X)} = r$. For a proof of the latter see [20], p480 (Also see Appendix B). □

**Lemma 5** (Lemma 2.1 in [21]). *Let $f \in \mathbb{F}[X]$ be a nonzero polynomial such that the degree of $f$ in $x_i$ is bounded by $r_i$, and let $S_i \subseteq \mathbb{F}$ be of size at least $r_i + 1$, for all $i \in [n]$. Then there exists $(s_1, s_2, \ldots, s_n) \in S_1 \times S_2 \times \ldots \times S_n$ with $f(s_1, s_2, \ldots, s_n) \neq 0$.*

## 3 Decomposing a $\pi$-Oblivious ABP into Independent Sets of Polynomials

We need the following lemma, or rather its corollary, for the construction of the generator for $\pi$-oblivious ABPs in Section 4. For any $f \not\equiv 0$ computed by a $\pi$-oblivious ABP, the lemma gives us a decomposition satisfying some useful independence properties. It will be sufficient to state the lemma for $\pi$ being the identity permutation.

**Lemma 6.** *let $k \geq 1$, and let $A$ be an oblivious ABP of width $w$ with source $s$ and sink $t$ having variable order $x_1, x_2, .., x_{2n}$. Suppose $\widehat{A} \not\equiv 0$. Then we can write for some $w' \leq w$, $f = \sum_{i \in [w']} f_i g_i$, where*

1. *$\{f_1, f_2, \ldots, f_{w'}\} \subseteq \mathbb{F}[x_1, x_2, \ldots, x_n]$ and $\{g_1, g_2, \ldots, g_{w'}\} \subseteq \mathbb{F}[x_{n+1}, x_{n+2}, \ldots, x_{2n}]$ are both independent sets of polynomials.*

2. *$\forall a \in \mathbb{F}^{w'}, \sum_{i \in [w']} a_i f_i$ can be computed by an oblivious ABP of width $w$ with variable order $x_1, x_2, .., x_n$.*

3. *$\forall a \in \mathbb{F}^{w'}, \sum_{i \in [w']} a_i g_i$ can be computed by an oblivious ABP of width $w$ with variable order $x_{n+1}, x_{n+2}, \ldots, x_{2n}$.*

*Proof.* Let $V$ be the set of variables used in $A$. Pick an arbitrary level $L$ of nodes $v_1, v_2, \ldots, v_w$ such that $V \cap \{x_1, x_2, \ldots, x_n\}$ appear on edges in layers before $L$, and $V \cap \{x_n, x_{n+1}, \ldots, x_{2n}\}$ appear on edges in layers after $L$. For $i \in [w]$, let $f_i = \widehat{A_{s,v_i}}$ and $g_i = \widehat{A_{v_i,t}}$. We proceed in two phases. First we arrange for a decomposition where the $f_i$s are independent. Then we will deal with the $g_i$s.

Wlog. assume that $f_1, \ldots, f_k$ is a maximum size independent set of polynomials. Since $f \not\equiv 0$, we know that not all $f_i \equiv 0$. So $k \geq 1$. For $j > 0$, any $f_{k+j}$ can be written as a linear combination of $f_1, \ldots, f_k$. Let $A'$ be an equivalent ABP obtained from $A$ as follows. First, $A'$ is just as $A$ from the source up to the level $L$, except that we drop $v_{k+1}, \ldots v_w$ from $L$. Let us use $L'$ to denote the modified level $L$. $L'$ is followed by a constant layer, where $f_1, \ldots, f_w$ are computed (relative to $s$). After this we attach all the levels of $A$, just as they followed $L$ in $A$. We have that $f = \sum_{i \in [k]} f_i g'_i$, where $f_i = \widehat{A'_{s,v_i}}$ and $g'_i = \widehat{A'_{v_i,t}}$. The $f_i$s satisfy the first two conditions of the lemma. The $g'_i$s are in $\mathbb{F}[x_{n+1}, x_{n+2}, \ldots, x_{2n}]$. This completes the first phase.

For the next phase, wlog. assume that $g'_1, \ldots, g'_l$ is a maximum size independent set. Say these correspond to nodes $w_1, \ldots, w_l$, respectively. That is, $\widehat{A'_{w_i,t}} = g'_i$. Since $f \not\equiv 0$, we know that $l \geq 1$. Symmetrically to the first phase, but now going in the direction from sink to source, we modify $A'$ into an equivalent ABP $A''$. $A''$ is the same as $A'$ from the sink back to the level $L'$, except that we drop nodes other than $w_1, \ldots, w_l$ from $L'$. Above this is a constant level, where we compute $g'_1, \ldots, g'_k$ (relative to the sink). Above this we attach all level from $A'$, just as they appear from $s$ to $L'$ in $A'$. We now have arranged that $f = \sum_{i \in [l]} f''_i g'_i$, where $f''_i = \widehat{A''_{s,w_i}}$ and $g'_i = A''_{w_i,t}$, for $i \in [l]$. Observe that for each $i \in [l]$, $f''_i = f'_i + \text{Linear}(f_{l+1}, \ldots, f_k)$. Hence $\{f''_1, \ldots, f''_l\}$ is an independent set of polynomials. All required properties of the lemma are now clearly satisfied. □

**Corollary 1.** *Let $k \geq 1, n \geq 3$ and let $1 < i < n$. Let $A$ be a read $r$ oblivious ABP, with source $s$ and sink $t$ having variable order $x_1, x_2, \ldots, x_{i-1}, x_i, x_{i+1}, \ldots, x_n$. We use $y$ as alias for $x_i$. Let $f = \partial \widehat{A}/\partial y$. Suppose $\widehat{A}$ depends on $y$, that is $f \not\equiv 0$. Then we can write for some $r' \leq r$, $f = \sum_{i \in [r']} p_i q_i$, where*

1. *$\{p_1, p_2, \ldots, p_{r'}\} \subseteq \mathbb{F}[x_1, x_2, \ldots, x_{i-1}]$ and $\{q_1, q_2, \ldots, q_{r'}\} \subseteq \mathbb{F}[x_{i+1}, x_{i+2}, \ldots, x_n]$ are both independent sets of polynomials.*

2. *$\forall a \in \mathbb{F}^{r'}$, $\sum_{i \in [r']} a_i p_i$ can be computed by a read $r$ oblivious ABP with variable order $x_1, x_2, \ldots, x_{i-1}$.*

3. *$\forall a \in \mathbb{F}^{r'}$, $\sum_{i \in [r']} a_i q_i$ can be computed by a read $r$ oblivious ABP with variable order $x_{i+1}, x_{i+2}, \ldots, x_n$.*

*Proof.* We make changes to $A$ by modifying the edges in the $y$-layer as follows: for a variable edge (labeled with $y$), label it with 1. For a constant edge, remove it. The resulting ABP $A'$ computes $f$. Then in the proof of Lemma 6, take the level $L$ to be the starting level of the original $y$-layer. As $|L|$ is bounded by the number of $y$-variables in the $y$-layer of $A$, we are done. □

## 4 A Generator for $\pi$-Oblivious ABPs

We assume $|\mathbb{F}|$ is large enough. The explicit requirement on $|\mathbb{F}|$ will become clear after the description of the generator. For now, let us fix $S = \{\alpha_1, \ldots, \alpha_N\} \subseteq \mathbb{F}$, for some $N$. Denote by $S_m = \{\alpha_1, \ldots, \alpha_m\}$, for $1 \leq m \leq N$. Let $Z = \{z_1, z_2, \ldots\}$, $Y = \{y_1, y_2, \ldots\}$, $U = \{u_1, u_2, \ldots\}$ and $V = \{v_1, v_2, \ldots\}$ be sets of variables. For $k \geq 1$, we use $Z_k$ to denote the $k$-tuple of variables $(z_1, z_2, \ldots, z_k)$, similarly for $Y_k$, $U_k$ and $V_k$. Define the function $\ell$ on natural numbers by $\ell(k,r) = 2rk + 1$. In the following, per abuse of notation, we write $(Z_{\ell(k,r)}, U_k, V_k)$ to denote the tuple $Z_{\ell(k,r)} \# U_k \# V_k$.

For every $k \geq 0$, $r \geq 1$ and a variable $w$, let $H^{k,r}(w) = (H^{k,r}_1(w), H^{k,r}_2(w), \ldots H^{k,r}_{\ell(k,r)+2k}(w))$, where for each $i \in [\ell(k,r) + 2k]$, $H^{k,r}_i$ is the $i$th Lagrange interpolation polynomial on the set $S_{\ell(k,r)+2k}$. $H^{k,r}_i$ is a univariate polynomial in $w$ of degree $\ell(k,r) + 2k - 1$, satisfying that $\forall \alpha_j \in S_{\ell(k,r)+2k}$, $H^{k,r}_i(\alpha_j) = 1$ if $i = j$ and 0 otherwise. For $k \geq 1$, and two variables $u$ and $v$, let $E^k(u,v) = (u \cdot L^k_1(v), \ldots, u \cdot L^k_{2^k}(v))$, in which $L^k_i$ is the $i$th Lagrange interpolation polynomial on the set $S_{2^k}$.

For $k \geq 0$ and $r \geq 1$, we define the polynomial mapping $F^{k,r}(Z_{\ell(k,r)}, U_k, V_k) : \mathbb{F}^{\ell(k,r)+2k} \to \mathbb{F}^{2^k}$ inductively as follows:

1. $F^{0,r}(z_1) = z_1$, and

2. For clarity we use $y_1, y_2, \ldots, y_{2r}$ as aliases for the variables $z_{\ell(k,r)+1}, z_{\ell(k,r)+2}, \ldots, z_{\ell(k,r)+2r}$, respectively. We take $F^{k+1,r}(Z_{\ell(k,r)}, Y_{2r}, U_{k+1}, V_{k+1})$ to be equal to the following $2^{k+1}$-tuple of polynomials:

$$E^{k+1}(u_{k+1}, v_{k+1}) + F^{k,r}(Z_{\ell(k,r)}, U_k, V_k) \# F^{k,r}\left((Z_{\ell(k,r)}, U_k, V_k) + T^{k,r}(Y_{2r})\right),$$

where $T^{k,r} : \mathbb{F}^{2r} \to \mathbb{F}^{\ell(k,r)+2k}$ is defined by $T^{k,r}(Y_{2r}) = \sum_{i \in [r]} y_i \cdot H^{k,r}(y_{r+i})$.

From the construction we can see that in order to accommodate for $S$, $|\mathbb{F}|$ should be no less than $\max(\ell(k,r) + 2k, 2^k)$. Now let us compute $F^{1,r}$ to get a sense and for later use. We obtain

$$\begin{aligned} F^{1,r} &= E^1(u_1, v_1) + F^{0,r}(z_1)\#F^{0,r}(z_1 + (z_{1+1} + \cdots + z_{1+r})) \\ &= (u_1 L_1^1(v_1) + z_1, u_1 L_2^1(v_1) + z_1 + \cdots + z_{1+r}). \end{aligned}$$

Note that $z_{2+r}, \ldots, z_{1+2r}$ are not used in the Lagrange interpolation in the $T^{0,r}$ part. By a straightforward induction, one can prove the following bound for the individual degree of a variable in $F^{k,r}$.

**Proposition 1.** $\forall k \geq 2$ and $r \geq 1$, the individual degree of any variable in any component of $F^{k,r}$ is at most $\prod_{j \in [k-1]} (\ell(j,r) + 2j)(\ell(j,r) + 2j - 1))$.

The main theorem of this section is proved next. It shows that the generator $F^{k,r}$ works for the class $\mathcal{C}$ of polynomials computed by read $r$ $\pi$-oblivious ABPs, where there is one single fixed order $\pi$ of the variables for the entire class $\mathcal{C}$. Wlog. the order is assumed to be $x_1, x_2, \ldots$. A generator for any other fixed order, is obtained by permuting the components of the output of the generator in the appropriate way. To make the algebraic geometry go through in the proof, we will assume that $\mathbb{F}$ is algebraically closed. We will remove this requirement subsequently with Corollary 2.

**Theorem 2.** Let $\mathbb{F}$ be an algebraically closed field. Let $k \geq 0$, and let $A$ be a $\pi$-oblivious ABP of read $r \geq 1$ with variable order $x_1, x_2, \ldots, x_{2^k}$. Suppose $A$ computes $f$, then $f \equiv 0 \iff f(F^{k,r}) \equiv 0$.

*Proof.* The "$\Rightarrow$"-direction is trivial, so it suffices to show that if $f \not\equiv 0$, then $f(F^{k,r}) \not\equiv 0$. We prove this by induction on $k$. For $k = 0$ it is obvious. For $k = 1$, we know there exists $(a, b)$ such that $f(a, b) \neq 0$. Recall $F^{1,r} = (u_1 L_1^1(v_1) + z_1, u_1 L_2^1(v_1) + z_1 + \cdots + z_{1+r})$. Then setting $c$ to be the assignment of $(Z_{2r+1}, u_1, v_1)$ as $z_1 = a$, $z_2 = b - a$ and other variables to 0, would give $f(F^{1,r}) = f(a, b) \neq 0$. So $f(F^{1,r}) \not\equiv 0$.

Now let $k \geq 1$. For the induction step from $k$ to $k+1$, we need to prove that $F^{k+1,r}$ works for an oblivious read $r$ ABP polynomial $f$ with variables $x_1, \ldots, x_{2^{k+1}}$. We use $X$ as an alias for $x_{2^k}$, and $\Lambda$ as an alias for $\alpha_{2^k}$. Let $g = \partial f / \partial X$, and note that $f = g \cdot X + f|_{X=0}$, since $f$ is multilinear. Wlog. we can assume that $f$ depends on $X$. Namely, since $f$ is multilinear, if $f$ does not depend on any variable, i.e. $\forall i, \partial f/\partial x_i \equiv 0$, then $f \in \mathbb{F}$ (even if $char(\mathbb{F}) > 0$). Clearly the theorem holds in this case. Otherwise, the rest of the proof goes through mutate mutandis by selecting $X$ to be the median variable (w.r.t. the variable order $x_1, x_2, \ldots$) of variables that $f$ depends on. Thus $g \not\equiv 0$. We claim that the following holds:

**Claim 1.** $h := g(F^{k+1,r})|_{v_{k+1}=\Lambda} \not\equiv 0$

Before proving Claim 1, let us show that this is sufficient to complete the proof of Theorem 2. Consider $f(F^{k+1,r})|_{v_{k+1}=\Lambda}$. It is equal to the following:

$$\begin{aligned} h \cdot \left(F_{2^k}^{k+1,r}|_{v_{k+1}=\Lambda}\right) + \left((f|_{X=0})(F^{k+1,r})\right)|_{v_{k+1}=\Lambda} &= \\ h \cdot \left((E_{2^k}^{k+1,r} + P(Z_{\ell(k,r)}, U_k, V_k))|_{v_{k+1}=\Lambda}\right) + \left((f|_{X=0})(F^{k+1,r})\right)|_{v_{k+1}=\Lambda} &= \\ h \cdot (u_{k+1} + P(Z_{\ell(k,r)}, U_k, V_k)) + \left((f|_{X=0})(F^{k+1,r})\right)|_{v_{k+1}=\Lambda}, \end{aligned}$$

for some polynomial $P$ in variables $(Z_{\ell(k,r)}, U_k, V_k)$. Observe that $\left((f|_{X=0})(F^{k+1,r})\right)|_{v_{k+1}=\Lambda}$ does not contain the variable $u_{k+1}$. The same holds for $P(Z_{\ell(k,r)}, U_k, V_k))$. Hence $h \cdot u_{k+1}$ cannot be canceled, and we conclude that $f(F^{k+1,r})|_{v_{k+1}=\Lambda} \not\equiv 0$. This implies $f(F^{k+1,r}) \not\equiv 0$.

We will now prove Claim 1. Let $F'^{k+1,r} = F^{k+1,r} - E^{k+1,r}$. Note that since $f$ is multilinear, $g$ does not depend on $X$. Hence $g(F^{k+1,r})|_{v_{k+1}=\Lambda} = g(F'^{k+1,r})$. We will show that $g(F'^{k+1,r}) \not\equiv 0$. We have that

$$F'^{k+1,r} = F^{k,r}(Z_{\ell(k,r)}, U_k, V_k)\#F^{k,r}\left((Z_{\ell(k,r)}, U_k, V_k) + T^{k,r}(Y_{2r})\right).$$

Again we will use $y_1, y_2, \ldots, y_{2r}$ as alias for the variables $z_{\ell(k,r)+1}, z_{\ell(k,r)+2}, \ldots, z_{\ell(k,r)+2r}$, respectively. Corollary 1 gives us that we can write $g = \sum_{i \in [r']} p_i q_i$, for some $r' \leq r$, where

1. $\{p_1, p_2, \ldots, p_{r'}\} \subseteq \mathbb{F}[x_1, x_2, \ldots, x_{2^k-1}]$ and $\{q_1, q_2, \ldots, q_{r'}\} \subseteq \mathbb{F}[x_{2^k+1}, x_{2^k+2}, \ldots, x_{2^{k+1}}]$ are both independent sets of polynomials.

2. $\forall a \in \mathbb{F}^{r'}$, $\sum_{i \in [r']} a_i p_i$ can be computed by an oblivious ABP of read $r$ with variable order $x_1, x_2, \ldots, x_{2^k-1}$.

3. $\forall a \in \mathbb{F}^{r'}$, $\sum_{i \in [r']} a_i q_i$ can be computed by an oblivious ABP of read $r$ with variable order $x_{2^k+1}, x_{2^k+2}, \ldots, x_{2^{k+1}}$.

For any $a \in \mathbb{F}^{r'}$ with $a \neq \overline{0}$, $\sum_{i \in [r']} a_i p_i \not\equiv 0$, and this sum can be computed by an oblivious ABP of read $r$ with variable order $x_1, x_2, \ldots, x_{2^k-1}$. Hence by induction hypothesis $\sum_{i \in [r']} a_i p_i(F^{k,r}) \not\equiv 0$. Let $\hat{p}_i = p_i(F^{k,r}(z_1, \ldots, z_{\ell(k,r)}, U_k, V_k))$. The above shows that $P := \{\hat{p}_1, \hat{p}_2, \ldots, \hat{p}_{r'}\}$ is an independent set of polynomials. Let $\hat{q}_i = q_i(F^{k,r}(z_1, \ldots, z_{\ell(k,r)}, U_k, V_k))$. Similarly we have that $Q := \{\hat{q}_1, \hat{q}_2, \ldots, \hat{q}_{r'}\}$ is an independent set of polynomials.

Since $\hat{p}_1 + \hat{p}_2 + \ldots + \hat{p}_{r'} \not\equiv 0$, there exists input $c \in \mathbb{F}^{\ell(k,r)+2k}$ so that if we let $a_i = \hat{p}_i(c)$, then $a = (a_1, a_2, \ldots, a_{r'}) \neq \overline{0}$. Let $V \subseteq \mathbb{F}^{\ell(k,r)+2k}$ be the algebraic set defined by the system of equations

$$\{\hat{p}_i(z_1, \ldots, z_{\ell(k,r)}, U_k, V_k) = a_i : \forall i \in [r']\}$$

We know this system has a solution namely $c$. Since $\mathbb{F}$ is assumed to be algebraically closed, by Exercise 1.9 p. 8 in [15] (See Appendix B), we know that each irreducible component of $V$ has dimension at least $\ell(k,r) + 2k - r'$. Since the system is solvable there must exist at least one irreducible component, and since $r \geq 1$, $\ell(k,r) + 2k - r' \geq 3$.

Let $W \subseteq \mathbb{F}^{\ell(k,r)+2k}$ be the algebraic set defined by the equation $\sum_{i \in [r']} a_i \hat{q}_i(z_1, \ldots, z_{\ell(k,r)}, U_k, V_k) = 0$. Since $Q$ is an independent set of polynomials the l.h.s. of the above equation is a nonzero polynomial. In case the l.h.s. is a non-zero constant, then we are done. Namely, letting $b \in \mathbb{F}^{\ell(k+1,r)+2(k+1)}$ be the assignment where we set $(z_1, \ldots, z_{\ell(k,r)}, U_k, V_k)$ to $c$, $y_1, \ldots, y_r$ to 0, and the remaining variables arbitrarily, would give $g(F'^{k+1,r})(b) = \sum_{i \in [r']} a_i \hat{q}_i(z_1, \ldots, z_{\ell(k,r)}, U_k, V_k)(b) \neq 0$. Otherwise, we know by Proposition 1.13 in [15] (See Appendix B), that $W$ is a finite union of hypersurfaces each of dimension $\ell(k,r)+2k-1$ (these correspond to the irreducible factors of $\sum_{i \in [r']} a_i \hat{q}_i(z_1, \ldots, z_{\ell(k,r)}, U_k, V_k)$ ). We want to argue that $V + \text{Im } T$ cannot be contained in $W$. Namely, to see the consequence, suppose we have $c' = c'' + T(d)$, for $c'' \in V$ and $d \in \mathbb{F}^{2r}$, with $c' \notin W$. Then letting $b \in \mathbb{F}^{\ell(k+1,r)+2k}$ be the assignment where we set $(z_1, \ldots, z_{\ell(k,r)}, U_k, V_k)$ to $c''$ and $Y_{2r} := d$ gives that $g(F'^{k+1,r})(b) = \sum_{i \in [r']} p_i(F^{k,r}(c''))q_i(F^{k,r}(c'' + T(d))) = \sum_{i \in [r']} \hat{p}_i(c'')\hat{q}_i(c'' + T(d)) = \sum_{i \in [r']} a_i \hat{q}_i(c') \neq 0$.

We complete the proof by showing that the Zariski-closure of $V + \text{Im } T$ has dimension greater than $\dim W$.

**Claim 2.** $\dim \overline{V + \text{Im } T} = \ell(k,r) + 2k$.

*Proof.* Observe that for any $r'' \leq r$, $\text{Im } T$ contains any $r''$-dimensional coordinate subspace of $\mathbb{F}^{\ell(k,r)+2k}$. Namely, by setting $y_{r+i} = \alpha_{j_i}$, for all $i \in [r]$, where $\alpha_{j_1}, \alpha_{j_2}, \ldots, \alpha_{j_r}$ are distinct elements of $S_{\ell(k,r)+2k}$, we obtain $\sum_{i \in [r]} y_i \cdot H^{k,r}(y_{r+i}) = \sum_{i \in [r]} y_i \cdot H^{k,r}(\alpha_{j_i}) = \sum_{i \in [r]} y_i \cdot e_{j_i}$, where $e_1, e_2, \ldots, e_{\ell(k,r)+2k}$ are standard basis vectors of $\mathbb{F}^{\ell(k,r)+2k}$. Hence the claim follows from Lemma 4. □

The above claim implies that $V + \text{Im } T \not\subset W$. By the above remarks, this gives that $g(F'^{k+1,r})(b) \neq 0$, for some $b$. This proves Claim 1, and finishes the proof of the theorem. □

**Corollary 2.** *Let $\mathbb{F}$ be any field. Let $k \geq 0$, and let $A$ be a $\pi$-oblivious ABP over $\mathbb{F}$ of read $r \geq 1$ with variable order $x_1, x_2, \ldots, x_{2^k}$. Suppose $A$ computes the polynomial $f \in \mathbb{F}[x_1, x_2, \ldots, x_{2^k}]$. Then in the construction of $F^{k,r}$ selecting any set $S$ of size $\max(\ell(k,r) + 2k, 2^k)$ contained in $\mathbb{F}$ (or an arbitrary field extension $\mathbb{G}$ of $\mathbb{F}$, if $\mathbb{F}$ is not large enough) yields that $f \equiv 0 \iff f(F^{k,r}) \equiv 0$,*

*Proof.* First consider the case when $\text{char}(\mathbb{F}) = 0$. In this case we take $S = \{0, 1, 2, \ldots\}$. Let $\bar{\mathbb{F}}$ be the algebraic closure of $\mathbb{F}$. Interpreting $A$ as an ABP over $\bar{\mathbb{F}}$, we can apply Theorem 2 to conclude $f \equiv 0 \iff f(F^{k,r}) \equiv 0$. All coefficients of $F^{k,r}$ are rational numbers and thus lie inside $\mathbb{F}$. Hence the property $f \equiv 0 \iff f(F^{k,r}) \equiv 0$ also holds when considering we work over $\mathbb{F}$.

In case $\text{char}(\mathbb{F}) > 0$, if $|\mathbb{F}|$ is not large enough, by allowing ourselves to use elements from the extension $\mathbb{G}$, we can still get the required $S$. Then similarly as above, by considering the algebraic closure of $\mathbb{G}$ and applying Theorem 2, the required generator property follows, considering one works over $\mathbb{G}$. □

## 5 A Black-Box PIT Algorithm for $\pi$-OABPs

---
**Algorithm 1** PIT Algorithm for read $r$ $\pi$-OABPs.

---
**Input:** Black-box access to $f \in \mathbb{F}[x_1, x_2, \ldots, x_n]$ computed by a $\pi$-OABP with read $r$.
**Output:** returns **true** iff $f \equiv 0$.
1: let $k$ be such that $2^{k-1} < n \leq 2^k$.
2: let $D = \prod_{j \in [k-1]} (\ell(j, r) + 2j)(\ell(j, r) + 2j - 1)$.
3: let $S_{D+1}$ be an arbitrary subset of $\mathbb{F}$ (or an extension field of $\mathbb{F}$ if $|\mathbb{F}| < D + 1$) of size $D + 1$.
4: let $R = S_{D+1}^{\ell(k,r)+2k}$.
5: compute $A = F^{k,r}(R)$.
6: permute the vectors in $A$ according to $\pi$.
7: **for all** $a \in A$ **do**
8:   check whether $f(a) = 0$.
9: **end for**
10: **return true** if in the previous stage no nonzero was found, **false** otherwise.

---

**Theorem 3.** *Let $\mathbb{F}$ be an arbitrary field. Using black-box Algorithm 1 we can check deterministically in time $2^{O(r \log r \cdot \log^2 n \log \log n)}$ whether a given polynomial $f \in \mathbb{F}[x_1, x_2, \ldots, x_n]$ computed by a read $r$ $\pi$-OABP is identically zero or not. If $\text{char}(\mathbb{F}) > 0$, the algorithm is granted black-box access to extension fields of $\mathbb{F}$.*

*Proof.* By Lemma 2, we can assume wlog. that $f$ is computed by a read $r$ $\pi$-oblivious ABP. By Theorem 2, we see that $f \equiv 0 \Leftrightarrow f(F^{k,r}) \equiv 0$. By Proposition 1, the individual degree of variables of $f(F^{k,r})$ can be bounded by $D = \prod_{j \in [k-1]} (\ell(j, r) + 2j)(\ell(j, r) + 2j - 1)$. Correctness now follows from Lemma 5. Bounding $D$ by $(2rk + 2k)^{2k}$, and knowing that the number of variables of $f(G^{k,r})$ is $2rk + 2k + 1$, the theorem follows by straightforward arithmetic.

We remark that the hitting set $A$, will be constructed over an extension field of $\mathbb{F}$ if $|\mathbb{F}| < \max(\ell(k, r) + 2k, 2^k)$ or $|\mathbb{F}| < D + 1$. In the former case, it is because of having enough interpolation points to define the generator. In the latter case it is in order to apply Lemma 5, as was done in the above. To work over the extension field the algorithm by Shoup [22] can be used to obtain an irreducible polynomial of degree $d$ over $\mathbb{F}$ in time $poly(d)$. With our demands, it suffices for the degree of this polynomial to be bounded by $O(\log n \log r + \log n \log \log n)$. Field operations in the extension field then take time $poly(\log n, \log r)$, assuming a unit cost model for operations in $\mathbb{F}$. The cost of constructing $A$ this way, can thus easily be seen to be subsumed by the time bound given in the theorem. □

We remark that the above implies that read $polylog(n)$ $\pi$-OABPs can be tested in $\text{DTIME}[2^{O(polylog(n))}]$.

## 6 Separation Results and Lower Bounds for OABPs

**Definition 4.** *For a given $\pi$-OABP $A$ with $X = \{x_1, \ldots, x_{2n+1}\}$, and two variable sets $Y = \{y_1, \ldots, y_n\}$ and $Z = \{z_1, \ldots, z_n\}$. We call the partition $B : X \setminus \{x_{\pi^{-1}(n+1)}\} \to Y \cup Z$ by mapping $x_{\pi^{-1}(i)}$ to $y_i$, and $x_{\pi^{-1}(n+1+i)}$ to $z_i$ for $i \in [n]$ the middle partition of $A$.*

By making use of the partial derivative matrix, we obtain a lower bound on the number of reads for an OABP as follows:

**Lemma 7.** *Given a $\pi$-OABP $A$ on variables $X = \{x_1, \ldots, x_{2n+1}\}$ of read $r$, let $f = \partial \widehat{A}/\partial x_{\pi^{-1}(n+1)}$. Let $B$ be the middle partition for $A$. Then we have that $r \geq \mathrm{rank}(M_{f^B})$.*

*Proof.* Due to Lemma 2, we can assume that $A$ is oblivious. Let the edges labeled with $x_{\pi^{-1}(n+1)}$ be $e_1$, $\ldots$, $e_{r'}$, for $r' \leq r$, and let $g_i = A_{\widehat{source(A),begin(e_i)}}$ and $h_i = A_{\widehat{end(e_i),sink(A)}}$, for $i \in [r']$. Note that taking the partial derivative w.r.t. $x_{\pi^{-1}(n+1)}$ is equivalent to setting $x_{\pi^{-1}(n+1)}$ to 1, and removing all constant labeled edges in the $x_{\pi^{-1}(n+1)}$-layer. Then we get that $f = \sum_{i \in [r']} g_i \cdot h_i$. From the definition of the middle partition, $g_i^B$ and $h_i^B$ are on two disjoint variable sets, respectively. Thus from the third property of Lemma 3, $\mathrm{rank}(M_{g_i^B \cdot h_i^B}) = 1$. So we have $\mathrm{rank}(M_{f^B}) = \mathrm{rank}(M_{(\sum_{i \in [r']} g_i \cdot h_i)^B}) = \mathrm{rank}(\sum_{i \in [r']} M_{g_i^B \cdot h_i^B}) \leq \sum_{i \in [r']} \mathrm{rank}(M_{g_i^B \cdot h_i^B}) \leq r' \leq r$. The claim follows. □

## 6.1 Exponential Lower Bounds for Permanent, Determinant and Full Rank Polynomials

**Theorem 4.** *Any OABP computing the permanent or determinant of an $n \times n$ matrix of variables has size $\Omega(2^n/n)$ and read $\Omega(2^n/n^2)$.*

*Proof.* Theorem 5 in [7] (which derives from [17]), states that in the model where edges are labeled with linear forms, any ordered ABP $A$ requires $\Omega(2^n)$ nodes to compute the permanent or determinant. The size lower bound then follows from Lemma 1. The lower bound on the number of reads follows from the fact that there must be $\Omega(2^n)$ edges in $A$, each having at least one variable. Hence some variable is read at least $\Omega(2^n/n^2)$ many times in $A$. We get the bound of the theorem, by observing that the conversion of Lemma 1 does not increase the read: if a variable $x_i$ is read $r$ times it will appear $r$ times in a linear form of the new created homogeneous ABP. □

Ryser's Formula [23] states the $n \times n$ permanent equals $\sum_{S \subset [n]} (-1)^{|S|} \prod_{i \in [n]} \sum_{j \notin S} x_{ij}$. Hence we see that the $n \times n$ permanent can be computed by an oblivious ABP of size $O(n^2 2^n)$ and read $2^n$. For the determinant, simply computing the sum of all $n!$ terms can be done by an oblivious ABP of both size and read $2^{O(n \log n)}$.

**Theorem 5.** *Any OABP $A$ over variables $X = \{x_1, \ldots, x_{2n+1}\}$, $\mathcal{W} = \{w_{i,j,k}\}_{i,j,k \in [2n+1]}$ using constants from $\mathbb{F}$ computing the polynomial $p$ constructed in the proof of Theorem 1 requires some variable to be read at least $2^n$ times.*

*Proof.* Interpret $A$ as working over the field $\mathbb{F}(\mathcal{W})$. The theorem then follows by combining Lemma 7 and Theorem 1. □

We can also interpret this to be giving a stronger lower bound (seen as a function of the number of variables), but for a polynomial which uses $O(n^3)$ transcendental constants in its definition.

**Corollary 3.** *For any field $\mathbb{F}$, and any extension field $\mathbb{G}$ of $\mathbb{F}$ of transcendence degree at least $(2n+1)^3$, there exists a polynomial $p \in \mathbb{G}[x_1, x_2, \ldots, x_{2n+1}]$, such that any OABP over $\mathbb{G}$ computing $p$ requires some variable to be read at least $2^n$ times.*

## 6.2 Separation of $R_{k-1}$-OABP and $R_k$-OABP

Consider the elementary symmetric polynomial of degree $k$ in $n$ variables given by $S_n^k = \sum_{S \subset [n], |S|=k} \prod_{i \in S} x_i$.

**Theorem 6.** *$S_n^k$ is an $R_k$-OABP-polynomial, but not a $R_{k-1}$-OABP-polynomial, for $n \geq 2k-1$, $k \geq 2$.*

To prove Theorem 6, it is enough to prove the following two lemmas:

**Lemma 8.** $S_n^k$ can not be computed by an $R_{k-1}$-OABP, for $n \geq 2k-1$, $k \geq 2$.

*Proof.* We only need to prove the statement for $n = 2k - 1$. This is because for $n > 2k - 1$, if there is an $R_{k-1}$-OABP $A$ computing $S_n^k(x_1, \ldots, x_n)$, setting edges labeled with $x_{2k}, \ldots, x_n$ to 0 in $A$ will give an $R_{k-1}$-OABP computing $S_{2k-1}^k(x_1, \ldots, x_{2k-1})$. So if $S_{2k-1}^k(x_1, \ldots, x_{2k-1})$ can not be computed by an $R_{k-1}$-OABP, it follows that $S_n^k(x_1, \ldots, x_n)$ can not be computed by an $R_{k-1}$-OABP, for any $n > 2k - 1$.

For $n = 2k - 1$, in order to get a contradiction, suppose there exists some $R_{k-1}$-OABP $A$ computing $S_n^k$. Since $S_n^k$ is a symmetric polynomial, we can assume the order of variables in $A$ is $(x_1, \ldots, x_n)$. Let $B$ be the middle partition for $A$, and let $wt(e)$ denote the Hamming weight of $e$. Note that $\partial S_n^k / \partial x_k = S_{n-1}^{k-1}(x_1, \ldots, x_{k-1}, x_{k+1}, \ldots, x_{2k-1})$. By Lemma 7, $\text{rank}(M_{(S_{n-1}^{k-1})_B}) \leq k - 1$. However, we have $\text{rank}(M_{(S_{n-1}^{k-1})_B}) \geq k$. This is because for $e, f \in \{0, 1\}^{k-1}$, $M_{(S_{n-1}^{k-1})_B}(e, f) = 1$ if and only if $wt(e) + wt(f) = k - 1$. If we let $E = \{e_0, e_1, \ldots, e_{k-1}\} \subseteq \{0, 1\}^{k-1}$ be an arbitrary set of vectors with weight $0, 1, 2, \ldots, k-1$, respectively, then the minor of $M_{(S_{n-1}^{k-1})_B}$ indexed by rows $E$ and columns $E$ is a permutation matrix. We have reached a contradiction. □

**Lemma 9.** $S_n^k$ can be computed by an $R_k$-OABP of size $O(kn)$, for $n \geq k \geq 1$.

*Proof.* For any $j \in [n - k + 1]$, let $s_j^k = S_{n-j+1}^k(x_j, \ldots x_n)$. We prove the following stronger statement by induction: for each $k$, there is an $R_k$-OABPs $A_k$ with variable order $x_1, \ldots, x_n$, such that for every $j \in [n - k + 1]$, there is a node $v_j$ with $(A_k)_{v_j, t} = s_j^k$, where $t = sink(A_k)$. Per abuse of notation, this node $v_j$ is denoted by $s_j^k$ as well. We will ensure that on any path from $s_j^k$ to $t$ only variables from $\{x_j, \ldots, x_n\}$ appear. Initially, we will not worry about layering the ABP.

For the base case $k = 1$. For any $j \in [n]$, $s_j^1 = x_j + \ldots + x_n$. So for $j < n$, $s_j^1 = x_j + s_{j+1}^1$. Hence the required read-once program $A_1$ is easily given. Next consider $k > 1$.

We will use the $R_{k-1}$-OABP for $A_{k-1}$ to get the $R_k$-OABP for $A_k$. Observe the following equation holds for any $1 \leq j < n - k + 1$,
$$s_j^k = x_j \cdot s_{j+1}^{k-1} + s_{j+1}^k, \tag{2}$$
and that $s_{n-k+1}^k = x_{n-k+1} \cdot s_{n-k+2}^{k-1}$.

The program $A_k$ is constructed as follows. First of all, the sink node will be equal to the old sink of $A_{k-1}$. For all $j \in [n - k + 1]$, we add a new node $s_j^k$. From $s_{n-k+1}^k$ we add a path to $s_{n-k+1}^{k-1}$ of weight $x_{n-k+1}$. Then using Equation (2), working from $j = n - k$ down to 1, we create a path from $s_j^k$ to $s_{j+1}^k$ of weight 1 and a path from $s_j^k$ to $s_{j+1}^{k-1}$ with weight $x_j$. The node $s_1^k$ can be taken as the new source, as the polynomial $s_1^k$ equals $S_n^k$. For clarification we have supplied some figures (Figures 1, 2 and 3). Since we maintain the invariant that on any path from $s_j^k$ to $t$ only variables from $\{x_j, \ldots, x_n\}$ appear, it is immediately clear that adding the path from $s_j^k$ with weight $x_j$ to $s_{j+1}^{k-1}$ leaves the program ordered w.r.t. the order $x_1, x_2, \ldots, x_n$. Furthermore, it is clear that the invariant is maintained. A similar comments holds regarding the path added from $s_j^k$ to $s_{j+1}^k$ with weight 1.

One can draw the ordered ABP on a grid, as has been show for example for $A_2$ in Figure 1. By considering the diagonals as levels on this grid, one can conclude we have given a leveled $\pi$-oblivious ABP with order $x_1, x_2, \ldots, x_n$ of size $O(kn)$ for $S_n^k$. □

### 6.3 Exponential Gap of Reads Between $\pi$-OABP and $\pi'$-OABP

In this section we show that under different permutations $\pi$ and $\pi'$, the gap between the number of reads for the models $\pi$-OABP and $\pi'$-OABP can be exponentially large.

**Theorem 7.** Given $X = \{x_0, x_1, \ldots, x_{2n-1}, x_{2n}\}$, $n \geq 1$, there exists a polynomial $p$ on $X$, and two permutations $\pi$ and $\pi'$ on $X$, such that 1) There exists a read-once $\pi$-OABP computing $p$, and 2) Any $\pi'$-OABP computing $p$ requires read $2^n$.

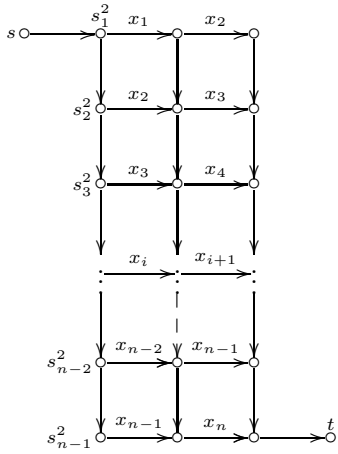

Figure 1: $R_2$-OABP for $S_n^2(x_1, \ldots x_n)$

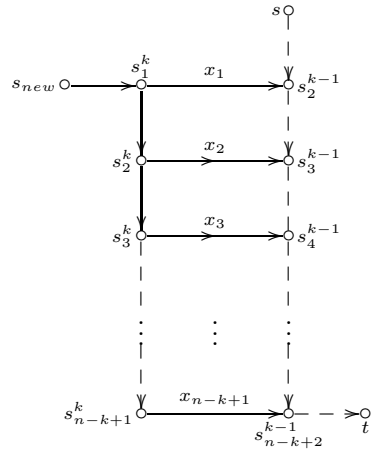

Figure 2: Induction Step : $R_k$-OABP for computing $S_n^k(x_1, \ldots, x_n)$

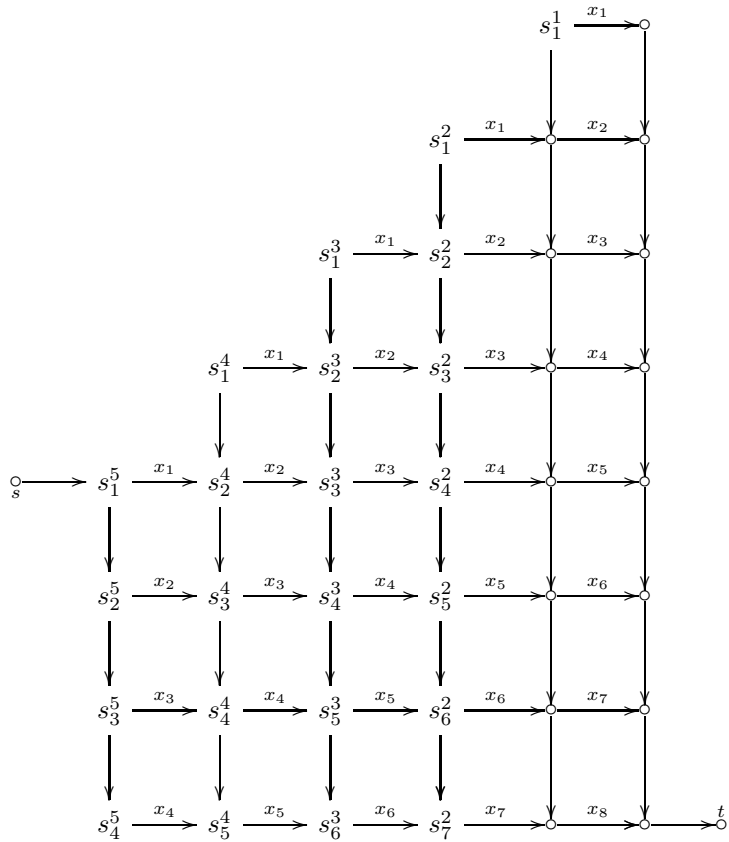

Figure 3: Example : $R_5$-OABP for $S_8^5(x_1, \ldots x_8)$ with redundant nodes still to be removed.

*Proof.* Define $p(x_0, x_1, \ldots, x_{2n}) = x_0 \prod_{i \in [n]}(x_{2i-1} + x_{2i} + x_{2i-1}x_{2i})$. Let $p_0 = x_0$, and $p_i = x_{2i-1} + x_{2i} + x_{2i-1}x_{2i}$, for $i \in [n]$. Let $\pi$ determine the variable order $x_0, x_1, \ldots, x_{2n}$. First, it is easy to construct an $R_1$-OABP $A_i$ computing $p_i$ with order $(x_{2i-1}, x_{2i})$, for $i \in [n]$, and there is an $R_1$-OABP $A_0$ computing $x_0$. To construct $\pi$-OABP $A$ computing $p$, we compose $A_i$ sequentially, by connecting $sink(A_{i-1})$ and $source(A_i)$, for $i \in [n]$.

Let $\pi'$ determine the variable order $x_1, x_3, \ldots, x_{2n-1}, x_0, x_2, \ldots, x_{2n}$. Let $A$ be a $\pi'$-OABP computing $p$, and suppose the read of $A$ is $r$. Let $q = \partial p/\partial x_0$, and let $B$ be the middle partition for $A$. By Lemma 7, $r \geq \text{rank}(M_{q^B}) = \text{rank}(M_{\prod_{i \in [n]} p_i^B}) = \prod_{i \in [n]} \text{rank}(M_{p_i^B}) = \prod_{i \in [n]} 2 = 2^n$. $\square$

# References

[1] J.T. Schwartz. Fast probabilistic algorithms for polynomial identities. *J. Assn. Comp. Mach.*, 27:701–717, 1980.

[2] R. Zippel. Probabilistic algorithms for sparse polynomials. In *Proceedings of the International Symposium on Symbolic and Algebraic Manipulation (EUROSAM '79)*, volume 72 of *Lect. Notes in Comp. Sci.*, pages 216–226. Springer Verlag, 1979.

[3] V. Kabanets and R. Impagliazzo. Derandomizing polynomial identity testing means proving circuit lower bounds. *Computational Complexity*, 13(1–2):1–44, 2004.

[4] M. Agrawal. Proving lower bounds via pseudo-random generators. In *Proc. 25th Annual Conference on Foundations of Software Technology and Theoretical Computer Science*, pages 92–105, 2005.

[5] M. Agrawal and V. Vinay. Arithmetic circuits: A chasm at depth four. In *Proc. 49th Annual IEEE Symposium on Foundations of Computer Science*, pages 67–75, 2008.

[6] N. Saxena. Progress of polynomial identity testing. Technical Report ECCC TR09-101, Electronic Colloquium in Computational Complexity, 2009.

[7] M. Jansen. Lower bounds for syntactically multilinear algebraic branching programs. In *Proc. 33rd International Symposium on Mathematical Foundations of Computer Science*, volume 5162 of *Lect. Notes in Comp. Sci.*, pages 407–418, 2008.

[8] R.E. Bryant. On the complexity of vlsi implementations and graph representations of boolean functions with application to integer multiplication. *IEEE Trans. Computers*, 40(2):205–213, 1991.

[9] R. Raz and A. Shpilka. Deterministic polynomial identity testing in non commutative models. *Computational Complexity*, 14(1):1–19, 2005.

[10] A. Shpilka and I. Volkovich. Improved polynomial identity testing of read-once formulas. In *Approximation, Randomization and Combinatorial Optimization. Algorithms and Techniques*, volume 5687 of *LNCS*, pages 700–713, 2009.

[11] M. Jansen, Y. Qiao, and J. Sarma M.N. Deterministic identity testing of read-once algebraic branching programs, 2009. http://arxiv.org/abs/0912.2565.

[12] M. Jansen. Weakening assumptions for deterministic subexponential time non-singular matrix completion, 2010. To Appear, 27th International Symposium on Theoretical Aspects of Computer Science (STACS 2010).

[13] R. Raz. Multilinear formulas for permanent and determinant are of super-polynomial size. *J. Assn. Comp. Mach.*, 56(2):1–17, 2009.

[14] L. Valiant. Completeness classes in algebra. Technical Report CSR-40-79, Dept. of Computer Science, University of Edinburgh, April 1979.

[15] R. Hartshorne. *Algebraic Geometry*. Graduate Texts in Mathematics, Vol 52. Springer Verlag, 1977.

[16] R. Impagliazzo, N. Nisan, and A. Wigderson. Pseudorandomness for network algorithms. In *Proc. 26th Annual ACM Symposium on the Theory of Computing*, pages 356–364, 1994.

[17] N. Nisan. Lower bounds for non-commutative computation: extended abstract. In *Proc. 23rd Annual ACM Symposium on the Theory of Computing*, pages 410–418, 1991.

[18] R. Raz. Separation of multilinear circuit and formula size. In *Proc. 45th Annual IEEE Symposium on Foundations of Computer Science*, pages 344–351, 2004.

[19] R. Raz and A. Yehudayoff. Balancing syntactically multilinear arithmetical circuits. *Computational Complexity*, 17(4):515–535, 2008.

[20] D. Cox, J. Little, and D. O'Shea. *Ideals, Varieties, and Algorithms, Second Edition*. Undergraduate Texts in Mathematics. Springer Verlag, 1996.

[21] N. Alon. Combinatorial nullstellensatz. *Combinatorics, Probability and Computing*, 8(1–2):7–29, 1999.

[22] V. Shoup. New algorithms for finding irreducible polynomials over finite fields. In *Proc. 29th Annual IEEE Symposium on Foundations of Computer Science*, pages 283–290, 1988.

[23] H.J. Ryser. *Combinatorial Mathematics*, volume 14 of *Carus Mathematical Monograph*. Mathematical Association of America, 1963.

# A  Proof of Theorem 1

Let $X = \{x_1, \ldots, x_{2n+1}\}$, $Y = \{y_1, \ldots, y_n\}$, $Z = \{z_1, \ldots, z_n\}$, and $W = \{w_{i,j,k}\}_{i,j,k \in [2n+1]}$. For a field $\mathbb{F}$, let $\mathbb{F}(W)$ be the field of rational functions of $\mathbb{F}$ in variables $W$.

In this section we define a polynomial $p(X) \in \mathbb{F}(W)[X]$, such that $\partial p/\partial x_i$ is full rank for every $i \in [2n+1]$, when working over $X \setminus \{x_i\}$. The construction extends the construction in Section 4.1 of [19]. We denote by $[i, j]$ the interval $\{k \in \mathbb{N} \mid i \leq k \leq j\}$. Let $|[i, j]|$ be the length of the interval, with $|[i, j]| = 0$ if $j < i$, and $|[i, j]| = j - i + 1$ if $i \leq j$. Let $X_{i,j}$ be the set of variables $x_m$, with $m \in [i, j]$. Let $W_{i,j}$ be the set of variables $w_{i,l,j}$, with $l \in [i, j]$. We are going to define a family of functions $f_{i,j}$, that is indexed by intervals $[i, j]$.

We first define $f_{i,j}$ for $[i, j]$'s with even length. For $[i, j]$ with length 0, $f_{i,j} := 1$. If $|[i, j]| > 0$, define $f_{i,j}$ as

$$f_{i,j} = (1 + x_i x_j) f_{i+1, j-1} + \sum_l w_{i,l,j} f_{i,l} f_{l+1,j},$$

where $l \in [i+1, j-2]$, with the constraint that $|[i, l]|$ is even.

For a partition $A$ on $X$, and an interval $[i, j]$ with length $2m$, it is proved in [19] that if $A$ is *balanced* on $[i, j]$, namely that the restriction of $A$ on $[i, j]$ assigns half of the variables to $Y$ and the other half to $Z$, then $\mathrm{rank}(M_{f_{i,j}^A})$ attains the maximum of $2^m$.

Now we define $f_{i,j}$ for $[i, j]$'s with odd length. For $[i, j]$ with length 1, $f_{i,i} := x_i$. If $|[i, j]| > 0$, define $f_{i,j}$ as

$$f_{i,j} = (1 + x_i x_j) f_{i+1, j-1} + \sum_{l \in [i, j-1]} w_{i,l,j} f_{i,l} f_{l+1,j}.$$

It is noted that the variables in $f_{i,j}$ are contained in $X_{i,j} \cup W_{i,j}$.

**Claim 3.** *Given interval $[i, j]$ of odd length $2\ell - 1$, for any $k \in [i, j]$ and a partition $A$ of $X$ which is balanced on $[i, j] \setminus \{k\}$, $M_{(\partial f_{i,j}/\partial x_k)^A}$ has rank $2^{\ell-1}$. In particular, $f := f_{1, 2n+1}$ satisfies that $\partial f/\partial x_k$ is a full rank polynomial for any $k \in [2n+1]$, when working over $X \setminus \{x_k\}$.*

*Proof.* We prove by induction on $\ell$, where $2\ell - 1$ is the length of the interval $[i, j]$. For $[i, j]$ of length 1 (i.e. $\ell = 1$), note that $f_{i,i} = x_i$ so it clearly satisfies the requirement. Next we do the induction step from $\ell$ to $\ell + 1$, so suppose $|[i,j]| = 2\ell + 1$. We distinguish between $k \in \{i, j\}$ and $k \in [i+1, j-1]$. We write $f'$ to indicate the derivative of $f$ w.r.t. $x_k$.

If $k \in \{i, j\}$, suppose $k = i$. Consider the term $w_{i,i,j} f_{i,i} f_{i+1,j} = w_{i,i,j} x_i f_{i+1,j}$. After taking the derivative w.r.t. $x_k$ it becomes $w_{i,i,j} f_{i+1,j}$. Since $f_{i+1,j}$ is a full rank polynomial, and $w_{i,i,j}$ does not appear in any other term of $f'_{i,j}$, we get that when $A$ is balanced on $[i,j] \setminus \{k\}$, $2^\ell \geq \text{rank}(M_{(f'_{i,j})^A}) \geq \text{rank}(M_{(f_{i+1,j})^A}) \geq 2^\ell$. The case when $k = j$ is dealt with by considering the term $w_{i,j-1,j} f_{i,j-1} f_{j,j}$.

Fix $k \in [i+1, j-1]$, we have that

$$f'_{i,j} = (1 + x_i x_j) f'_{i+1, j-1} + \sum_{j-1 \geq l \geq k} w_{i,l,j} f'_{i,l} f_{l+1,j} + \sum_{i \leq l < k} w_{i,l,j} f_{i,l} f'_{l+1,j},$$

since $x_k$ does not appear in $f_{l+1,j}$, for $l \geq k$, nor $f_{i,l}$, for $l < k$. To prove $f'_{i,j}$ is full rank when $A$ is balanced on $[i,j] \setminus \{k\}$, like in [19], we distinguish between two cases.

**Case one:** for every $l \in [i+1, j-1]$, such that $[i, l] \setminus \{k\}$ is of even length, $A$ is not balanced on $[i, l] \setminus \{k\}$. Wlog. assume $x_i$ is partitioned in $Y$, then as $A$ is not balanced on $[i,l] \setminus \{k\}$ for every $l \in [i+1, j-1]$, but balanced on $[i,j] \setminus \{k\}$, $x_j$ must be partitioned in $Z$. Thus $A$ is balanced on $[i+1, j-1] \setminus \{k\}$. So by induction hypothesis we get that $\text{rank}(M_{(f'_{i+1,j-1})^A}) = 2^{\ell-1}$, and note that $\text{rank}(M_{((1+x_i x_j)^A)}) = 2$. Thus $2^\ell \geq \text{rank}(M_{(f'_{i,j})^A}) \geq \text{rank}(M_{(f'_{i+1,j-1})^A}) \text{rank}(M_{(1+x_i x_j)^A}) = 2^\ell$, which yields $\text{rank}(M_{(f'_{i,j})^A}) = 2^\ell$.

**Case two:** there exists $l' \in [i+1, j-1]$, such that $[i, l'] \setminus \{k\}$ is of even length, and $A$ is balanced on $[i, l'] \setminus \{k\}$. Suppose $l' < k$, and consider the term $w_{i,l',j} f_{i,l'} f'_{l'+1,j}$. Note that $w_{i,l',j}$ does not appear in other terms. Thus $2^\ell \geq \text{rank}(M_{(f'_{i,j})^A}) \geq \text{rank}(M_{(f_{i,l'} f'_{l'+1,j})^A}) = \text{rank}(M_{(f_{i,l'})^A}) \text{rank}(M_{(f'_{l'+1,j})^A}) = 2^{|[i,l']|/2} 2^{|[l'+1,j]\setminus\{k\}|/2} = 2^\ell$, which gives us $\text{rank}(M_{(f'_{i,j})^A}) = 2^\ell$. If $l' \geq k$, by considering the term $w_{i,l',j} f'_{i,l'} f_{l'+1,j}$, we can get the same result. □

## B More Algebraic Geometry

We collect the theorems used in this article, omitting their proofs. Let $\mathbb{F}$ be an algebraically closed field. For a set of polynomials $S \subset \mathbb{F}[x_1, \ldots, x_n]$ we denote by $Z(S)$ the algebraic set of common zeroes of $S$. For a subset $V \subset \mathbb{F}^n$ we denote by $I(V)$ the *ideal* of $V$. $I(V)$ is the subset of polynomials in $\mathbb{F}[x_1, \ldots, x_n]$ that vanish on $V$. Taking complements of algebraic sets in $\mathbb{F}^n$ to be *open sets*, gives rise to a topology on $\mathbb{F}^n$, the so-called *Zariski* topology. For any set $S \subseteq \mathbb{F}^n$, $\overline{S}$ denotes the closure of $S$ in the Zariski topology. $\overline{S}$ equals the intersection of all algebraic sets containing $S$. If $\overline{S} = \mathbb{F}^n$, then $S$ is said to be *dense* in $\mathbb{F}^n$. A non-empty set $V$ of $\mathbb{F}^n$ is called *irreducible*, if it cannot be written as a union $V_1 \cup V_2$ of proper closed subsets of $V$. A *variety* is any algebraic sets which is irreducible. We remark that [20] calls any algebraic set a variety. We will keep to the terminology of [15].

**Proposition 2** (Corollary 4, p. 479 in [20])**.** *Let $V \subseteq \mathbb{F}^n$ be an algebraic set. Then the dimension of $V$ is equal to the largest integer $r$ for which there exist $r$ variables $x_{i_1}, \ldots, x_{i_r}$, such that $I(V) \cap \mathbb{F}[x_{i_1}, \ldots, x_{i_r}] = \{0\}$.*

A corollary of Proposition 2, which is not explicitly stated but given as a remark in [20], is the following:

**Corollary 4** (Variation of Proposition 5, p. 480 in [20])**.** *Let $V \subseteq \mathbb{F}^n$ be an algebraic set. Then the dimension of $V$ is the largest dimension of a coordinate subspace $H \subseteq \mathbb{F}^n$ for which a projection of $V$ onto $H$ is Zariski dense. That is, the closure of the projection of $V$ onto $H$ is $H$ itself.*

**Proposition 3** (Exercise 1.9, p. 8 in [15])**.** *Let $I \subseteq \mathbb{F}[x_1, \ldots, x_n]$ be an ideal that can be generated by $r$ elements. Then every irreducible component of $Z(I)$ has dimension $\geq n - r$.*

**Proposition 4** (Proposition 1.13, p. 7 in [15])**.** *A variety $Y$ in $\mathbb{F}^n$ has dimension $n - 1$ if and only if it is the zero set $Z(f)$ of a single non constant irreducible polynomial in $\mathbb{F}[x_1, \ldots, x_n]$.*